\documentclass[%
 reprint,
 amsmath,amssymb,
 aps,
    pra,
floatfix,
nofootinbib,
]{revtex4-2}

\usepackage{bm}
\usepackage{graphicx}
\usepackage{dcolumn}
\usepackage{amsmath}
\allowdisplaybreaks  
\usepackage[T1]{fontenc}
\usepackage{url}
\usepackage{booktabs}
\usepackage{amsfonts}
\usepackage{nicefrac}
\usepackage{microtype}
\usepackage{graphicx}
\usepackage{amssymb}
\usepackage{pifont}
\usepackage{wrapfig}
\usepackage{amssymb}
\usepackage{physics}
\usepackage{tikz}
\usepackage{enumitem}
\usepackage{makecell}
\usepackage{newtxtext}
\usepackage{nameref}
\usepackage{mathtools}
\usepackage{hyperref}
\usepackage{cleveref}
\usepackage{letltxmacro}
\usepackage{babel}
\usepackage{titlesec}
\renewcommand{\selectlanguage}[1]{}

\let\oldsubsection\subsection
\renewcommand{\subsection}[1]{
    \oldsubsection{#1}
    \vspace{-0.2em} 
}
\let\oldsubsubsection\subsubsection
\renewcommand{\subsubsection}[1]{
    \oldsubsubsection{#1}
    \vspace{-0.2em} 
}

\newcommand{\eqrefp}[1]{(Eq.~\ref{#1})}
\renewcommand{\eqref}[1]{Eq.~\ref{#1}}
\newcommand{\Hp}[2]{\mathcal{H}^{#2}_{#1}} %

\newcommand{\R}{\mathbb{R}}
\newcommand{\C}{\mathbb{C}}

\begin{document}
\preprint{APS/123-QED}

\title{General Oscillator-Based Ising Machine Models \texorpdfstring{\\}{:} with Phase-Amplitude Dynamics and Polynomial Interactions}

\author{Lianlong Sun}
\email{Contact author: lianlongsun@rochester.edu}
\author{Matthew X. Burns}%
\email{Contact author: mburns13@ur.rochester.edu}
\author{Michael C. Huang}%
\email{Contact author: michael.huang@rochester.edu}
\affiliation{Department of Electrical and Computer Engineering, University of Rochester, USA}

\begin{abstract}
\vspace{5pt}
We present an oscillator model with both phase and amplitude dynamics for oscillator-based Ising machines (OIMs). The model targets combinatorial optimization problems with polynomial cost functions of arbitrary order and addresses fundamental limitations of previous OIM models through a mathematically rigorous formulation with a well-defined energy function and corresponding dynamics. The model demonstrates monotonic energy decrease and reliable convergence to low-energy states. Empirical evaluations on 3-SAT problems show significant performance improvements over existing phase-amplitude models. Additionally, we propose a flexible, generalizable framework for designing higher-order oscillator interactions, from which we derive a practical method for oscillator binarization without compromising performance. This work strengthens both the theoretical foundation and practical applicability of oscillator-based Ising machines for complex optimization problems.

\end{abstract}
\maketitle

\section{INTRODUCTION}

The limitations of traditional von Neumann architectures in efficiently tackling combinatorial optimization problems have spurred the exploration of alternative computing paradigms. Among these, Ising machines, inspired by the Ising model from statistical mechanics, have emerged as a promising approach. The versatility of Ising machines extends across a diverse range of applications, including portfolio optimization \cite{tanahashi2019application, parizy2022cardinality}, drug discovery \cite{mao2023chemical}, vehicle routing \cite{bao2023ising, tsuyumine2024optimization, azad2022solving}, fraud detection \cite{stein2023exploring}, clustering \cite{matsumoto2022distance}, power grid optimization \cite{kirihara2023exploring}, wireless communication \cite{singh2022ising, sreedhara2023mu}, image restoration \cite{6621708} and machine learning \cite{niazi2024training,wuextending,song2024ds,instatrain,liuexpressive}. Additionally, Ising machines leverage multiple specialized hardware substrates, including quantum~\cite{johnson2011quantum,king2023quantum}, optical~\cite{clements2017gaussian,pierangeli2019large,marandi2014network,wu2025monolithically,calvanese2021all,wang2013coherent,cen2022large,gao2024photonic,pierangeli2020noise,sun2022quadrature}, CMOS~\cite{takemoto20192, matsubara2020digital, belletti2008janus, baity2014janus, yamaoka201520k, afoakwa_brim_2021, su2024flexspin, xie2022ising} and other emerging hardware platforms~\cite{honjo2021100, litvinenko2023spinwave, yue2024scalable,chen2024oscillatory,maher2024cmos,si2024energy,chen2025off}. By mapping the decision variables of an optimization problem onto interacting spins \cite{lucas_ising_2014}, Ising machines can efficiently search for solutions by allowing the system to naturally evolve towards its lowest energy state. This process corresponds to finding a configuration of spins that minimizes the system's energy, which directly translates to a high-quality solution to the original optimization problem. This approach offers the potential to significantly accelerate the solution of complex optimization problems that are intractable for classical computers. 

A major challenge in Ising machine implementations is that the original Ising model supports only up to second-order interactions. Mapping practical problems such as 3-SAT introduces third-order terms into the Hamiltonian, creating a fundamental mismatch between the problem structure and the physical implementation. A widely adopted approach to address this issue is cubic-to-quadratic reduction \cite{kolmogorov_what_2004, freedman_energy_2005, anthony_quadratic_2017, cilasun_3sat_2024}, which introduces auxiliary variables. However, this method often complicates the energy landscape, making it more difficult to navigate \cite{dobrynin_energy_2024}, and consequently leads to suboptimal performance \cite{bybee_efficient_2023, sharma_augmenting_2023}. Recent advances \cite{bybee_efficient_2023, kanao_simulated_2023, sharma_augmenting_2023, bashar_designing_2023} have proposed native supports for higher-order interactions, avoiding the need for any reductions. 

Ising machines based on networks implementing the Kuramoto model \cite{kuramoto_self-entrainment_1975} of synchronization have gained significant traction. These oscillator-based Ising machines (OIMs) \cite{wang_oim_2019, bashar_experimental_2020, dutta_understanding_2020, mallick_using_2020, ahmed_probabilistic_2020, chou_analog_2019, graber2024firefly, sreedhara2024novel, garg2024phase, ochs2021ising, sreedhara2023digital, cilasun2025coupled} based on Kuramoto model exploit the emergent collective behavior of synchronized oscillators to emulate spin interactions, with the system's natural tendency towards stable, synchronized states corresponding to low-energy solutions of the Ising Hamiltonian. While Kuramoto-based OIMs have shown promise, oscillator models that incorporate amplitude dynamics in addition to phase 
\cite{bybee_efficient_2023, matheny2019exotic, slavin2009nonlinear, gambuzza2016amplitude, garg2024improved} have aimed to provide a more complete physical description of the underlying hardware. Research has revealed that phase-amplitude formulations for nanoelectromechanical (NEMS) oscillators can diverge significantly from simplified phase-only descriptions under specific operating regimes \cite{matheny2019exotic}. Moreover, incorporating amplitude dynamics into spintronic oscillator models (Slavin's model~\cite{slavin2009nonlinear}) yields a more faithful representation of coupled magnetization processes and the system demonstrates improved time complexity over the Goemans-Williamson algorithm \cite{garg2024improved}. For consistency with \cite{bybee_efficient_2023}, we refer to models that integrate both amplitude and phase in the context of Ising machines as the \textit{Hopf oscillator model}.

Our study is motivated by the need for a faithful, general oscillator model that incorporates both phase and amplitude dynamics while supporting higher-order interactions. We begin by identifying critical limitations in the previously proposed Hopf oscillator model for optimization problems \cite{bybee_efficient_2023}:

\begin{itemize}[itemsep=0pt, leftmargin=1.5em]
\item \textbf{Lack of a Physically Meaningful Hamiltonian Definition:} A central concept in Ising machines is the system's energy, which should correspond to the objective function being optimized. In previous work, the defined energy function is generally complex-valued, lacking a clear physical interpretation. This contrasts with traditional Ising machines where energy is a real-valued function directly related to the optimization problem cost function.

\item \textbf{Non-Monotonic Energy Decrease:} Ideally, the system's energy should be able to decrease monotonically in the absence of introduced noise. In prior model~\cite{bybee_efficient_2023}, we observed that neither individual parts (real or imaginary) nor absolute value of the energy consistently decrease during the optimization process. 
\end{itemize}

To address these limitations, we have developed an oscillator model with complex-valued oscillator representation, based on Andronov-Hopf dynamics~\cite{nikonov2015coupled}, providing a rigorous framework for oscillator-based Ising machines. Our key contributions include:

\begin{itemize}[itemsep=0pt, leftmargin=1.5em]
    \item \textbf{Energy Definition:} We introduce the complex conjugate of the oscillator representation variable $ z $ to define a real-valued and physically meaningful energy function that directly corresponds to the objective function of the optimization problem.
    \item \textbf{Oscillator Dynamics Derivation:} Using Wirtinger calculus \cite{kreutz-delgado_complex_2009}, we derive the oscillator dynamics from our redefined energy function. These dynamics guarantee a monotonic decrease in energy under ideal conditions as the system evolves towards a solution.
    \item \textbf{Generalization to Higher-Order Interactions:} Our formulation naturally extends higher-order interactions among oscillators, thus broadening its applicability to more complex optimization problems.
    \item \textbf{Alternative Binarization Approach:} We propose an alternative method for binarizing the oscillator variable $z$, derived from the model formulation. This approach maintains solution quality while providing flexibility in implementation.
\end{itemize}

This rigorous framework not only provides a deeper theoretical understanding of OIMs but also leads to significant empirical performance improvements compared to existing phase-amplitude approaches. Remarkably, the energy function terms in our model closely resemble those found in nonlinear optical or four-wave mixing Hamiltonian \cite{Kanao_2021, quinn2024coherentisingmachinebased}. This connection suggests exciting new paths for realizing powerful Ising machines, potentially leveraging existing photonic technologies. Furthermore, we demonstrate that our model reduces to a previously proposed high-order, Kuramoto-based formulation~\cite{bashar_designing_2023} once amplitude dynamics are removed. Our work therefore generalizes previous polynomial frameworks, offering perspective on the relationship between Hopf and Kuramoto oscillator models for Ising machines. 

Our findings offer a stronger theoretical foundation for OIMs while providing actionable guidelines for leveraging their potential to address complex optimization challenges.

\section{BACKGROUND AND PRELIMINARIES}

\subsection{Ising Hamiltonian for Combinatorial Optimization}
1-D spin glass systems were originally proposed as a model for quenched disorder and frustration in metallic alloys~\cite{edwards_theory_1975, sherrington_solvable_1975}. For $N$ 1-D spins $s=(s_1,...,s_N)$ with a coupling matrix $J\in \mathbb{R}^{N\times N}$ and external bias $h\in \mathbb{R}^{N}$, the Ising spin glass (ISG) Hamiltonian, named after Ernst Ising for his earlier work on purely ferromagnetic systems~\cite{ising1925beitrag}, is given by
\begin{equation}
    H(s)=-\sum_{i<j}J_{ij}s_is_j-\sum_{i=1}^N h_is_i.
\end{equation}

Later work revealed that finding the ground state of a spin glass is NP-hard~\cite{barahona_computational_1982-2,istrailStatisticalMechanicsThreedimensionality2000}. Accordingly, a number of high-impact combinatorial problems can be mapped to an equivalent spin glass system~\cite{lucas_ising_2014, mohseni_ising_2022-1}. In particular, an ISG Hamiltonian can be trivially mapped to Quadratic Unconstrained Binary Optimization (QUBO) problems, a well-studied class within operations research~\cite{beasley_or-library_1990}.

The relative simplicity of the ISG Hamiltonian makes it highly amenable for hardware acceleration, leading to the advent of ``Ising machines'' (IMs)~\cite{mohseni_ising_2022-1}: devices which ``implement'' the Ising Hamiltonian in some manner. IM proposals include purely digital accelerators~\cite{aramon_physics-inspired_2019,goto_high-performance_2021}, analog devices evolving a continuous relaxation of the Ising Hamiltonian~\cite{wang_oim_2019,afoakwa_brim_2021,rajak_quantum_2023}, and hybrid digital/analog systems~\cite{inagaki_coherent_2016,hizzani_memristor-based_2024}.

\subsection{Solving 3-SAT problems with Ising Machine}
\label{sec:1_solving}
Ising machines have been successfully applied to quadratic unconstrained binary optimization (QUBO) problems~\cite{afoakwa_brim_2021,wang_oim_2019,honjo_100000-spin_2021-1} and even polynomial unconstrained binary optimization (PUBO)~\cite{bashar_designing_2023,sharma_combining_2023,bybee_efficient_2023} problems with the help of higher-order interactions. A generic Ising machine workflow usually involves \ding{172} converting the problem to a discrete Ising formula, \ding{173} mapping the formula to a continuous-time, dynamical Ising machine capable of seeking the minimum energy state and \ding{174} extracting the solution from the final device state. In this subsection, we demonstrate how to map 3-SAT problems to an Ising Machine as a preliminary step.

Consider a Boolean formula $f: \{0,1\}^N \to \{0,1\}$ in 3-literal conjunctive normal form (3-CNF)
\begin{equation}
    f(x) = \bigwedge_{j=1}^m(\ell_{j,1}\lor\ell_{j,2}\lor\ell_{j,3}),
\end{equation}
where each $\ell_{j,1}$ is a \emph{literal} (either a variable $x_p$ or its negation $\overline{x}_p$) and each disjunction of literals $(\ell_{j,1}\lor\ell_{j,2}\lor\ell_{j,3})$ is a \emph{clause}. The 3-SAT problem consists of finding some $x$ such that $f(x)=1$, i.e., an assignment satisfying all of the clauses. We restrict ourselves to 3-SAT without loss of generality, as any $K$-SAT problem can be reduced to an equivalent 3-SAT problem by adding $O(K\cdot m)$ variables and clauses~\cite{sipser_introduction_2012}.

Let $s=\{s_i\}_{i=1}^n\in \{-1,+1\}^n$ be the vector of Ising spins. We follow the standard convention that $s_i=1\iff x_i=1$ and $s_i=-1\iff x_i=0$~\cite{lucas_ising_2014}.

For each literal $\ell_{j,k}\in \{x_p, \overline{x}_p\}$ we define a spin function $L_{j,k}:\{-1,1\}^n\to \{0,1\}$ as
\begin{align}\label{eqn:mapping}
L_{j,k}(s) = \frac{1}{2}\begin{cases}
    1 - s_p & \text{ if } \ell_{j,k} = x_p\\
    1 + s_p & \text{ if } \ell_{j,k} = \overline{x}_p
\end{cases},
\end{align}
where $L_{j,k}(s)=0$ if the literal is satisfied, and $1$ otherwise.

A clause $c_j = (\ell_{j,1} \lor \ell_{j,2} \lor \ell_{j,3})$  is satisfied if at least one of its literals is true. Similarly to individual literals, we can define a spin function $C_j(s)$ as
\begin{align}
C_j(s) = L_{j,1}(s) \cdot L_{j,2}(s) \cdot L_{j,3}(s),
\end{align}
where $C_j(s)=0$ if $c_j$ is satisfied and 1 otherwise.

Substituting \eqref{eqn:mapping}, we can see that each $C_j$ is a spin polynomial of degree three
\begin{equation}
C_j(s) = \left(\frac{1 \pm s_{p}}{2}\right) \left(\frac{1 \pm s_{q}}{2}\right) \left(\frac{1 \pm s_{r}}{2}\right),
\label{eqn:clause}
\end{equation}
where $s_p$, $s_q$, and $s_r$ are spins corresponding to the variables of each literal.

We then define the objective function $E:\{-1,1\}^N\to \mathbb{Z}_+$ as the sum of all clause functions $C_j(s)$,
\begin{align}
E(s) = \sum_{j} C_j(s) = \sum_{j} L_{j,1}(s) L_{j,2}(s) L_{j,3}(s).
\end{align}
From \eqref{eqn:clause}, when a clause is satisfied, its corresponding $C_j$ evaluates to zero, contributing nothing to the total energy. In contrast, if a clause is unsatisfied, $C_j$ contributes a positive value of $1$ to $\mathcal{H}$. Therefore, minimizing $\mathcal{H}$ effectively minimizes the number of unsatisfied clauses. The optimization problem can thus be formally stated as
\begin{equation}\label{eqn:optimization_target}
\begin{aligned}
E &= \sum_{i=1}^{n} h_i s_i + \sum_{i=1}^{n} \sum_{j=1}^{n} J_{ij} s_i s_j \\ 
&+ \sum_{i=1}^{n} \sum_{j=1}^{n} \sum_{k=1}^{n} P_{ijk} s_i s_j s_k + \text{constant},
\end{aligned}
\end{equation}
where ${h}$, ${J}$, and ${P}$ are the coefficient vector, the coupling matrix, and the three-body interaction tensor, respectively. Minimizing $E$ corresponds to finding an assignment of spin variables $s_i$ that satisfies the maximum number of clauses. A solution with $E = 0$ indicates that all clauses are satisfied, thereby providing a solution to the original 3-SAT problem. Appendix{~\ref{appdx:sat}} provides a concrete example by mapping a single 3-SAT clause to the PUBO Hamiltonian.

In practice, analog Ising machines often represent binary spins $s$ by some physical quantities, thus relaxing the discrete phase space into a continuous one. We assume native device support for 3-body interactions and focus on the performance of the specific formulation. 3-body interactions can be implemented in analog electronic systems by means of multi-input gates (see~\cite{bashar_designing_2023,sharma_augmenting_2023} for more details).

\subsection{Kuramoto Oscillator Model}
The Kuramoto model~\cite{kuramoto_self-entrainment_1975} was originally proposed as a phase reduction method to study synchronization phenomena in weakly-coupled oscillator networks~\cite{gherardini_spontaneous_2018, pietras_network_2019}. A system of $N$ coupled oscillators with phase $\theta=(\theta_1, ..., \theta_N)$ is described by a system of coupled ODEs
\begin{equation}
    \frac{d\theta_i}{dt} = \kappa \sum_{\substack{j=1 \\ j \neq i}}^N J_{ij}\sin(\theta_i - \theta_j) + \omega_i,
\label{eqn:kuramoto_eq}
\end{equation}
where $\kappa$ and $J_{ij}$ describe global and pairwise interaction strengths respectively, and $\omega_i$ is the natural frequency of the $i^{th}$ oscillator. We set $\omega_i=0$ in \eqref{eqn:kuramoto_eq} assuming that all oscillators have identical $\omega_i$ and viewing the system in a rotating reference frame. Using spin glass interaction coefficients for $J$ has led to oscillator-based accelerators for combinatorial optimization: \emph{oscillator Ising machines} (OIM)~\cite{wang_oim_2019, bashar_experimental_2020, dutta_understanding_2020, mallick_using_2020, ahmed_probabilistic_2020, chou_analog_2019, graber2024firefly, sreedhara2024novel, garg2024phase, ochs2021ising, sreedhara2023digital,vaidya_creating_2022,zhang_review_2024}. 

We can trivially derive the global Lyapunov function associated with \eqref{eqn:kuramoto_eq} as
\begin{equation}
    E(\theta) = \kappa \sum_{i < j} J_{ij}\cos(\theta_i - \theta_j).
\end{equation}
The coupled oscillator system tends to minimize energy $E(\theta)$ over time \cite{lyapunov1992general}, laying the foundation for OIMs that tackle combinatorial optimization problems. The original OIM architecture~\cite{wang_oim_2019} utilized electronic LC-tank oscillators, and more recent proposals have broadened the choice of substrate to ring oscillators~\cite{lo_ising_2023,moy_1968-node_2022}, phase transition nano-oscillators~\cite{maher_cmos-compatible_2024, maher_highly_2024,dutta_ising_2021}, and magnetic tunnel junctions~\cite{grimaldi_evaluating_2023,si_energy-efficient_2024}. 

\subsection{Sub-harmonic Injection Locking}
The key idea of an OIM (and analog Ising machines in general) is to evolve the system towards a lower energy state, then map the state back to ISG spins. We term the mapping from oscillator's XY spin state (0 to 2$\pi$ in phase) to Ising spins (0 or $\pi$) as \emph{binarization}. However, the stationary points of the oscillator system may not have a known binarization. For instance, a two-oscillator system $H(\theta)=\cos(\theta_1-\theta_2)$ has infinitely many stationary points of the form
\begin{equation}
    \theta_1 -\theta_2=\pi.
\end{equation}

For a toy 2-oscillator system, the binarization is clear. However, coupled Kuramoto oscillator networks may not permit a closed-form binarization scheme. Kuramoto synchronization results in the oscillators bifurcating to aligned/non-aligned phases along some axis in the imaginary plane; however, that axis is not known a priori. To provide a simple, bijective mapping between oscillator fixed points and spin states, techniques such as sub-harmonic injection locking (SHIL) are introduced.

To provide simple binarization, previous work~\cite{wang_oim_2019} injects a sub-harmonic injection locking (SHIL) term into their spin dynamics,
\begin{equation}\label{eqn:kuramoto}
    \frac{d\theta_i}{dt} = \kappa \sum_{i<j}^NJ_{ij}\sin(\theta_i - \theta_j) - \kappa_{S}(t)\overbrace{\sin(2\theta_i)}^{\text{SHIL}}.
\end{equation}
The second harmonic creates a potential well at $\theta_i=n\pi$ for $n\in \mathbb{N}$, inducing sub-harmonic injection locking (SHIL). During annealing, periodic application of SHIL forces the oscillators to bifurcate, with appropriate hyperparameters ensuring a final two-cluster bifurcation.

\section{METHODS}

\subsection{Hopf Oscillator Model}
\label{sec:2_theoretical_analysis}
While the Kuramoto model is sufficient to study phase-only phenomena, real-world oscillators will have variability in both phase and amplitude. To enable a more faithful model, we can generalize from the phase-reduced representation to complex $z\in \mathbb{C}^N$. An oscillator model for combinatorial optimization and sampling is the Adronov-Hopf model (or ``Hopf model'' for brevity)~\cite{bybee_efficient_2023, nikonov2015coupled}. The evolution of each Hopf oscillator is given by a system of coupled ODEs
\begin{equation}\label{eqn:overall_model}
    \frac{dz_i}{dt}=f(z_i) + g(z),
\end{equation}

\noindent
where $ z_i $ describes the phase and amplitude of the $ i $-th oscillator, $f(z_i)$ represents the local dynamics of the oscillator, and $g(z)$ represents the coupling interaction between oscillators. In previous work~\cite{bybee_efficient_2023} (which is also the baseline we compare with), the local dynamics $f(z_i)$ is defined as
\begin{equation}
    f(z_i) = (\lambda_i + i\omega_i)z_i + \rho_i z_i|z_i|^2,
\end{equation}
where the center frequency is set to $\omega_i = 0$ for all oscillators. The parameters $\lambda_i$ and $\rho_i$ are chosen such that the system exhibits a stable limit cycle with unit amplitude. The term $g(z)$ is derived from the model energy function $\mathcal{H}$. Previous work~\cite{bybee_efficient_2023} directly maps the optimization problem to the system energy function $\mathcal{H}$ by replacing the Ising spin variable $s_i$ in \eqref{eqn:optimization_target} with the complex oscillator representation $z_i$ as follows:
\begin{equation}\label{eqn:stupid_hopf}
\begin{aligned}
\mathcal{H} 
&= \sum_{i=1}^{n} h_i z_i + \sum_{i=1}^{n} \sum_{j=1}^{n} J_{ij} z_i z_j \\
&+ \sum_{i=1}^{n} \sum_{j=1}^{n} \sum_{k=1}^{n} P_{ijk} z_i z_j z_k + \text{constant}.
\end{aligned}
\end{equation}
Several observations motivate us to investigate a new formulation:
\begin{enumerate}
    \item The energy $\mathcal{H}(z)$ is generally complex-valued. While the function is holomorphic and mathematically differentiable, the physical interpretation of a complex energy remains ambiguous.
    \item The model does not ensure that the real part, imaginary part, or absolute value of $\mathcal{H}(z)$ decreases monotonically under ideal conditions. This raises concerns about convergence and optimization efficiency. Lyapunov stability analysis, typically used to demonstrate that a system seeks a local minimum of an energy function, is not applicable here. 

\end{enumerate}

To address the challenges in accurately describing the system and the oscillator dynamics with a complex variable $z_i$, 

\begin{enumerate}
    \item We introduce the complex conjugate \(z_i^*\) into the energy function. This approach enables the formulation of a real-valued energy function \(\mathcal{H}'(z)\) that provides a meaningful physical interpretation of the system.
    \item 
    Using the new energy function, we derive system dynamics using gradient of the energy function. The resulting gradient system naturally drives the state toward minima of the real-valued energy function.
\end{enumerate}

Specifically, one of the energy functions of our Hopf oscillator model is defined as follows:
\begin{equation}\label{eqn:improved_hopf}
\begin{aligned}
\mathcal{H'} &= \frac{1}{2} \sum_{i=1}^{n} h_i (z_i + z_i^*) + \sum_{i=1}^{n} \sum_{j=1}^{n} J_{ij} z_i z_j^* \\
&+ \frac{1}{2} (\sum_{i=1}^{n} \sum_{j=1}^{n} \sum_{k=1}^{n} P_{ijk} z_i z_j z_k^* \\
&+ \sum_{i=1}^{n} \sum_{j=1}^{n} \sum_{k=1}^{n} P_{ijk} z_i z_j^* z_k^*) 
+ \text{constant}.
\end{aligned}
\end{equation}
Note that this construction is not unique; in Sec.~\ref{sec:general} we discuss a general framework for generating appropriate energy functions. 

With the introduction of the complex conjugate, the energy function is no longer holomorphic. However, by applying Wirtinger calculus \cite{kreutz-delgado_complex_2009}, we can rigorously derive the system dynamics. The core idea of Wirtinger calculus is to treat the complex function $f(z)$ as a function of both $z$ and its conjugate $z^*$ as independent variables. For non-holomorphic functions, this requires differentiation with respect to both variables. Appendix{~\ref{appdx:wirtinger}} contains a brief overview of Wirtinger calculus.

The symmetry of our energy functions with respect to $z$ and $z^*$ permits a simplification: we can compute the gradient by considering only the derivative with respect to either $z$ or $z^*$. Now we have, 
\begin{equation}
\begin{aligned}
\frac{\partial\mathcal{H'}}{\partial z_i} &= \frac{1}{2} h_i + \sum_{j=1}^{n} J_{ij} z_j^* \\
&+ \frac{1}{2} \left(\sum_{j=1}^{n} \sum_{k=1}^{n} P_{ijk}  z_j z_k^*  +\sum_{i=1}^{n} \sum_{k=1}^{n} P_{ijk}  z_i z_k^* \right.\\
&+ \left.\sum_{j=1}^{n} \sum_{k=1}^{n} P_{ijk} z_j^* z_k^*\right).
\end{aligned}
\end{equation}

For convenience, the coefficient $P$ is constructed to be symmetric under all permutations of its indices. It follows that
$
\sum_{j=1}^{n}\sum_{k=1}^{n}P_{ijk}\,z_{j}\,z_{k}^{*}
=
\sum_{i=1}^{n}\sum_{k=1}^{n}P_{ijk}\,z_{i}\,z_{k}^{*}
$. Therefore, summing over any two of the three indices yields the same result. To drive the system towards minimizing the energy, we set
\begin{equation}\label{eqn:grad}
\frac{dz_i}{dt} = -\left(\frac{\partial \mathcal{H'}}{\partial z_i}\right)^*.
\end{equation}
Furthermore, we have the identity
\begin{equation}\label{eqn:zzz}
    \frac{dz_i}{dt}=\left(\frac{dz_i^*}{dt}\right)^*,
\end{equation}
which follows from the definition of $z$ and $z^*$. Additionally, we have
\begin{equation}\label{eqn:crcr}
    \left(\frac{\partial \mathcal{H'}}{\partial z_i}\right)^* = \pdv{\mathcal{H'}}{z_i^*},
\end{equation}
which follows from the conjugation property of Wirtinger derivatives (see Appendix~\ref{appdx:wirtinger}). Combining these results with~\eqref{eqn:grad}, we obtain the total derivative:
\begin{equation}\label{eqn:improved_dynamics}
\begin{aligned}
\frac{d\mathcal{H'}}{dt} 
&= \sum_i\frac{\partial \mathcal{H'}}{\partial z_i}\frac{dz_i}{dt} + \sum_i\frac{\partial \mathcal{H'}}{\partial z_i^*}\frac{dz_i^*}{dt} \\
&= -2\sum_i\frac{\partial \mathcal{H'}}{\partial z_i}\left(\frac{\partial \mathcal{H'}}{\partial z_i}\right)^* \\
&= -2\sum_i \left[ \Re\left(\frac{\partial\mathcal{H'}}{\partial z_i}\right)^2 + \Im\left(\frac{\partial\mathcal{H'}}{\partial z_i}\right)^2\right] \leq 0,
\end{aligned}
\end{equation}
with equality if and only if $\frac{d \mathcal{H'}}{d z_i} = 0$, i.e., at a stationary point. Since the Hessian operator remains continuous in Wirtinger calculus~\cite{kreutz-delgado_complex_2009}, the dynamics \eqrefp{eqn:improved_dynamics} are locally Lyapunov stable within some neighborhood for all minima. Therefore,~\eqref{eqn:improved_hopf} is a suitable Lyapunov function for complex-domain Ising machines. In contrast, the energy formulation proposed in previous work~\cite{bybee_efficient_2023} is complex-valued and cannot serve as a viable Lyapunov function\footnote{Disregarding the physical interpretation of the complex-valued $\mathcal{H}$, we examine the behavior of the baseline model{~\cite{bybee_efficient_2023}} on a simple 3-variable problem, $E = s_1 s_2 s_3$, where $s_i \in \{-1, +1\}$. Both real and imaginary components exhibit oscillatory behavior, which leads to unstable dynamics. This can be demonstrated mathematically and empirically, and the observation extends to general cases.}.

\subsection{Compatibility with Kuramoto model}
\label{sec:3_bridge}
The Kuramoto model was originally derived as a phase reduction for networks of weakly coupled oscillators~\cite{kuramoto_self-entrainment_1975}. Both the phase-only model (including the quadratic Kuramoto model) and the Hopf model only approximate real-world oscillators. However, the Hopf model is more expressive, as it accounts for amplitude variations. One would expect that in the case of amplitude homogeneity, the Hopf model reduces to the phase-only model. Here, we show that indeed the Hamiltonian of our model, \eqref{eqn:improved_hopf} can be used to derive phase-only models found in previous literature in the context of cubic forms. In so doing, we demonstrate our model \ding{172} conforms to physical intuition and \ding{173} generalizes previous work.

Expressing $z_i=|z_i|e^{-i\theta_i}$, we have the equivalent expression
\begin{equation}\label{eqn:improved_hopf_phase_1}
\begin{aligned}
\mathcal{H'}(z) &= \frac{1}{2} \sum_{i=1}^{n} h_i |z_i|(e^{i\theta_i} + e^{-i\theta_i}) \\
&+ \sum_{i=1}^{n} \sum_{j=1}^{n} J_{ij} |z_i| |z_j|e^{i(\theta_i-\theta_j)} \\
&+ \frac{1}{2} (\sum_{i=1}^{n} \sum_{j=1}^{n} \sum_{k=1}^{n} P_{ijk} |z_i| |z_j| |z_k|e^{i(\theta_i+\theta_j-\theta_k)} \\
&+ \sum_{i=1}^{n} \sum_{j=1}^{n} \sum_{k=1}^{n} P_{ijk} |z_i| |z_j| |z_k|e^{-i(\theta_i+\theta_j-\theta_k)}) \\
&+ \text{constant}.
\end{aligned}
\end{equation}
Since $\mathcal{H'}(z)$ is real by construction, we have
\begin{equation}\label{eqn:improved_hopf_phase_2}
\begin{aligned}
\mathcal{H'}(z) &= \operatorname{Re}[\mathcal{H'}(z)]\\
&=\sum_{i=1}^{n}h_i |z_i|\cos(\theta_i) \\
&+ \sum_{i=1}^{n} \sum_{j=1}^{n} J_{ij} |z_i| |z_j|\cos(\theta_i-\theta_j) \\
&+ \sum_{i=1}^{n} \sum_{j=1}^{n} \sum_{k=1}^{n} P_{ijk} |z_i| |z_j| |z_k|\cos(\theta_i+\theta_j-\theta_k) \\
&+ \text{constant}.
\end{aligned}
\end{equation}
If we further assume amplitude homogeneity ($|z_i|=|z_j|=|z_k|=1$), which will remove the amplitude dynamics from the model, then we obtain
\begin{equation}\label{eqn:improved_hopf_phase_3}
\begin{aligned}
\mathcal{H'}(z) &= \sum_{i=1}^{n}h_i\cos(\theta_i) + \sum_{i=1}^{n} \sum_{j=1}^{n} J_{ij} \cos(\theta_i-\theta_j) \\
&+ \sum_{i=1}^{n} \sum_{j=1}^{n} \sum_{k=1}^{n} P_{ijk} \cos(\theta_i+\theta_j-\theta_k) 
+ \text{constant},
\end{aligned}
\end{equation}
which is identical to the form proposed in Ref.~\cite{bashar_designing_2023}.

\section{Results}

\subsection{Model Evaluation}

We compare our proposed Hamiltonian (\eqref{eqn:improved_hopf}) against the previously established baseline \cite{bybee_efficient_2023} (\eqref{eqn:stupid_hopf}), which also incorporates the amplitude dynamics for oscillator-based Ising machines. We simulate these models by numerically solving their governing differential equations.

\textbf{Overall Performance:}  We evaluate model performance on 1,700 uniform random 3-SAT benchmarks from SATLIB~\cite{hoos_satlib_2000}. Each set, \texttt{ufN-M}, denotes instances with N variables and M clauses; we use \texttt{uf20-91}, \texttt{uf50-218}, \texttt{uf75-325}, \texttt{uf100-430}, and \texttt{uf150-645}, all at a clause-to-variable ratio of approximately 4.3. This ratio lies at the empirical \textit{phase transition} in satisfiability, where random 3-SAT instances shift from mostly satisfiable to mostly unsatisfiable and where difficulty peaks \cite{monasson1999determining}. We define a \emph{solvable instance} as one where at least one of 100 independent runs (with different initial conditions) reaches zero energy within a sufficient simulation time ($T=136$), allowing both models to achieve their best performance on the benchmark.

\begin{figure}[ht]    \includegraphics[width=1\linewidth]{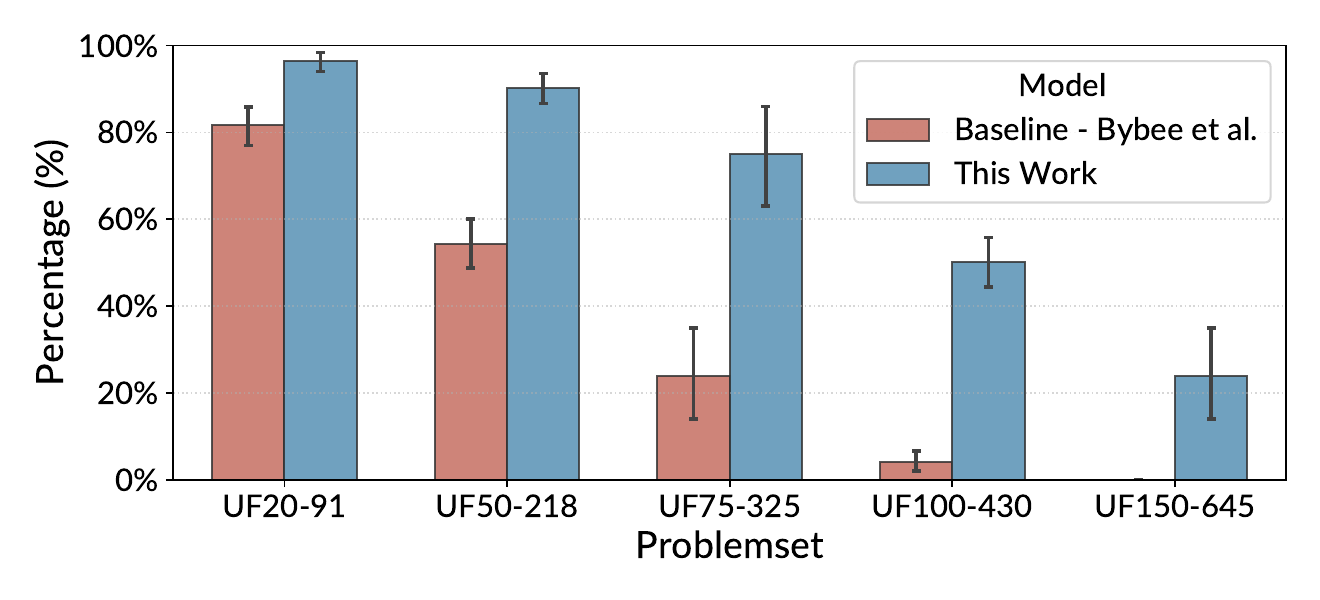}
    \caption{Percentage of solvable instances across problem sets (UF20-91 to UF150-645) from SATLIB \cite{hoos_satlib_2000} under 100 runs per instance. 
    Instances are classified as \emph{solvable} if at least one run reaches the ground state ($\mathcal{H}=0$). The error bars denote the 99\% bootstrap confidence intervals (computed over 10,000 resamples).
    The proposed Hopf model consistently outperforms the baseline model \cite{bybee_efficient_2023}.}
\label{fig:solvable}
\end{figure}
\begin{figure}[bht]
\includegraphics[width=1\linewidth]{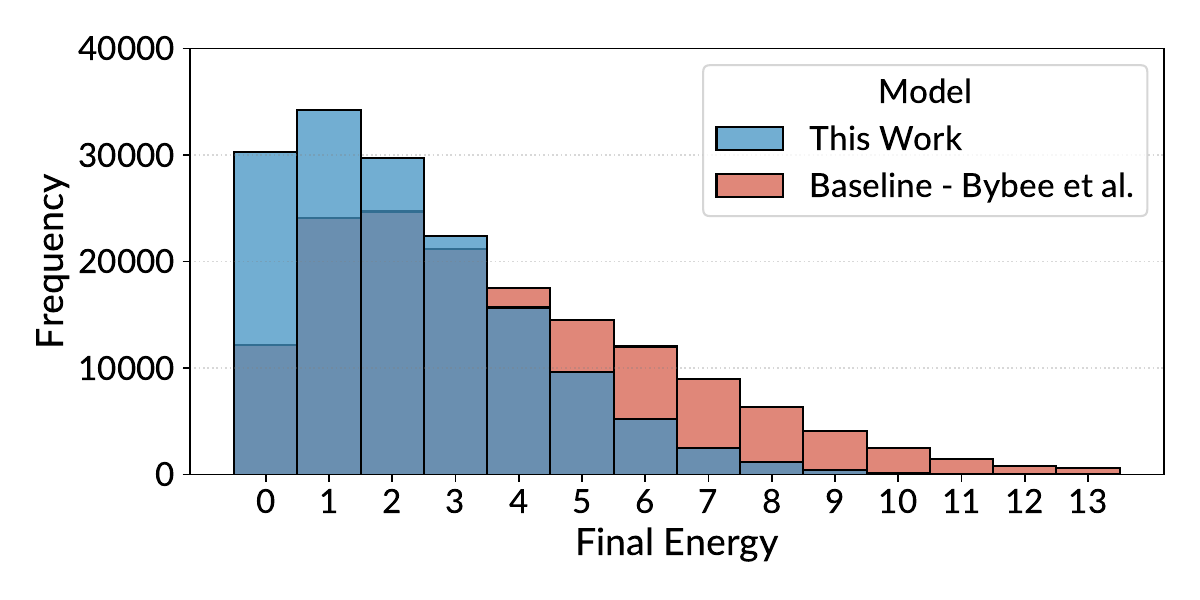}
    \includegraphics[width=1\linewidth]{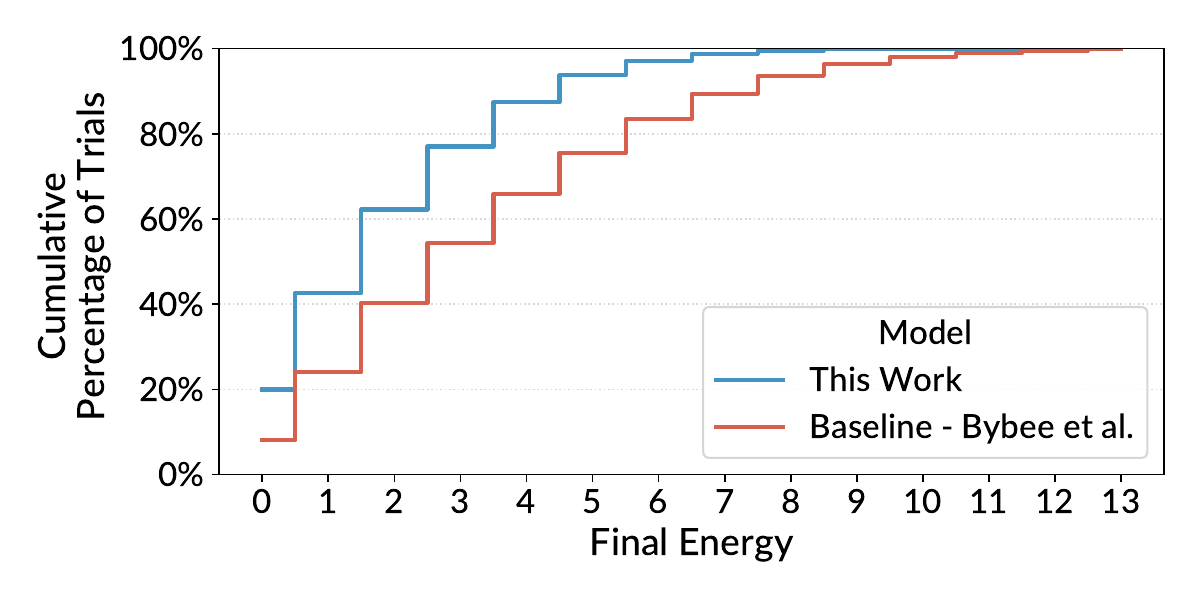}
    \caption{Final energy distribution for solving 3-SAT problems across approximately 170,000 runs per model. The proposed Hopf model achieves lower final energies with higher frequency, as shown in the histogram (top). The cumulative plot (bottom) further demonstrates the improved performance, showing a higher percentage of trials for the proposed model successfully reaching optimal or near-optimal solutions.}
\label{fig:energy_distribution}
\end{figure}

The following results show that our approach consistently outperforms the baseline Hopf model: We solve a higher percentage of instances and achieve lower final energy values more frequently. These improvements indicate superior optimization performance, as illustrated in Fig.~\ref{fig:solvable} and Fig.~\ref{fig:energy_distribution}.

\subsection{Higher-order Models with Wirtinger Potentials}
\label{sec:general}
For effective optimization of combinatorial problems and accurate representation of the Ising formulation, the energy function of an oscillator system ideally satisfies two key conditions:
\begin{enumerate}
    \item It should establish a clear correspondence between the continuous system energy and the original optimization problem. One approach to achieve this is through a real-valued energy function.
    
    \item Upon binarization, the energy function would optimally yield the same values as the discrete Ising formulation. For instance, with a complex oscillator representation $z$, one may map $\operatorname{Re}(z_i) > 0$ to $s_i = +1$ and $\operatorname{Re}(z_i) < 0$ to $s_i = -1$; alternatively, in a phase-only model, upon binarization, mapping $\phi_i = 0$ to $s_i = +1$ and $\phi_i = \pi$ to $s_i = -1$ achieves this equivalence.
\end{enumerate}

There are multiple ways to construct an energy function that satisfies these conditions. The most natural approach is to replace spin $s_i$ with $\frac{z_i+z_i^{*}}{2|z_i|}$. The resulting cubic terms of an energy function (ignoring the constant factor) will be

\begin{equation}
    \Hp{3}{C}(z) = \sum_{i,j,k} \left( z_i + z_i^{*} \right) \left( z_j + z_j^{*} \right) \left( z_k + z_k^{*} \right).
    \label{eqn:h0_3}
\end{equation}

We refer to this as the \emph{Complete potential} (or C-potential for short) for cubic interactions, as it involves a complete set of cross terms. However, a family of systematic subsets of its terms can each satisfy the desired conditions. Taking cubic terms as an example: 

\begin{equation}
\begin{aligned}
    \Hp{3}{0}(z)&= \sum_{i,j,k} \left( z_i z_j z_k + z_i^{*} z_j^{*} z_k^{*} \right),\\
    \Hp{3}{1}(z) &= \sum_{i,j,k} \left( z_i z_j z_k^{*} + z_i^{*} z_j^{*} z_k \right), \\
    &...
\end{aligned}
    \label{eqn:h0_1}
\end{equation} are all manifestly real and fulfill the required conditions. We number this family of cost functions $\{\mathcal{H}^0_{k},\mathcal{H}^1_{k},...,\mathcal{H}^{\lfloor k/2\rfloor}_{k}\}$ and refer to them collectively as \emph{Wirtinger potentials} for $\mathcal{H}$, with $\Hp{k}{p}$ referred to as the  $p$-potential of order $k$ for $\mathcal{H}$. Here $p$ refers to the number of conjugate terms in the leading product of each conjugate pair. For instance, each pair of $\Hp{4}{1}(z)$ has the form $z_iz_jz_kz_\ell^*+z_i^*z_j^*z_k^*z_\ell$, while $\Hp{4}{2}(z)$ conjugate pairs have the form $z_iz_jz_k^*z_\ell^*+z_i^*z_j^*z_kz_\ell$. A general definition for Wirtinger potentials of arbitrary order can be found in Appendix{~\ref{appdx:wirtinger_advanced}}. Each $\Hp{k}{p}$ is real-valued and has identical values for $z\in \C^n$, preserving Property 1. So too are properly scaled linear combinations of $\Hp{k}{p}$, including the special case $\Hp{k}{C}=\frac{1}{\lfloor k/2\rfloor}\sum_p \Hp{k}{p}$.

Wirtinger potentials enable flexibility in modeling oscillator systems while maintaining consistency with the Ising paradigm. For example, in a 3-SAT problem, the optimization target comprises three types of terms: linear, quadratic, and cubic (corresponding to $k = 1$, $2$, and $3$, respectively). This means that multiple oscillator modeling options exist for each type of term. The energy function {\eqref{eqn:improved_hopf}} is one such option, and our empirical evaluations indicate that this combination delivers the best performance in our benchmarks. However, it may not be the optimal choice for all problem types. In the following section ({\ref{sec:4_shil}}), we discuss a real case that leverages the variety of these models to achieve effective binarization for oscillator-based Ising machines. A thorough analysis of higher-order systems would require extensive theoretical and experimental study, which is beyond the scope of this work. Future research may explore these formulations to further enhance the capabilities of oscillator-based Ising machines.

\subsection{Binarization Behavior: Quadratic Problems}
\label{sec:4_shil}

\begin{figure}[h]
    \centering
    \includegraphics[width=0.75\linewidth]{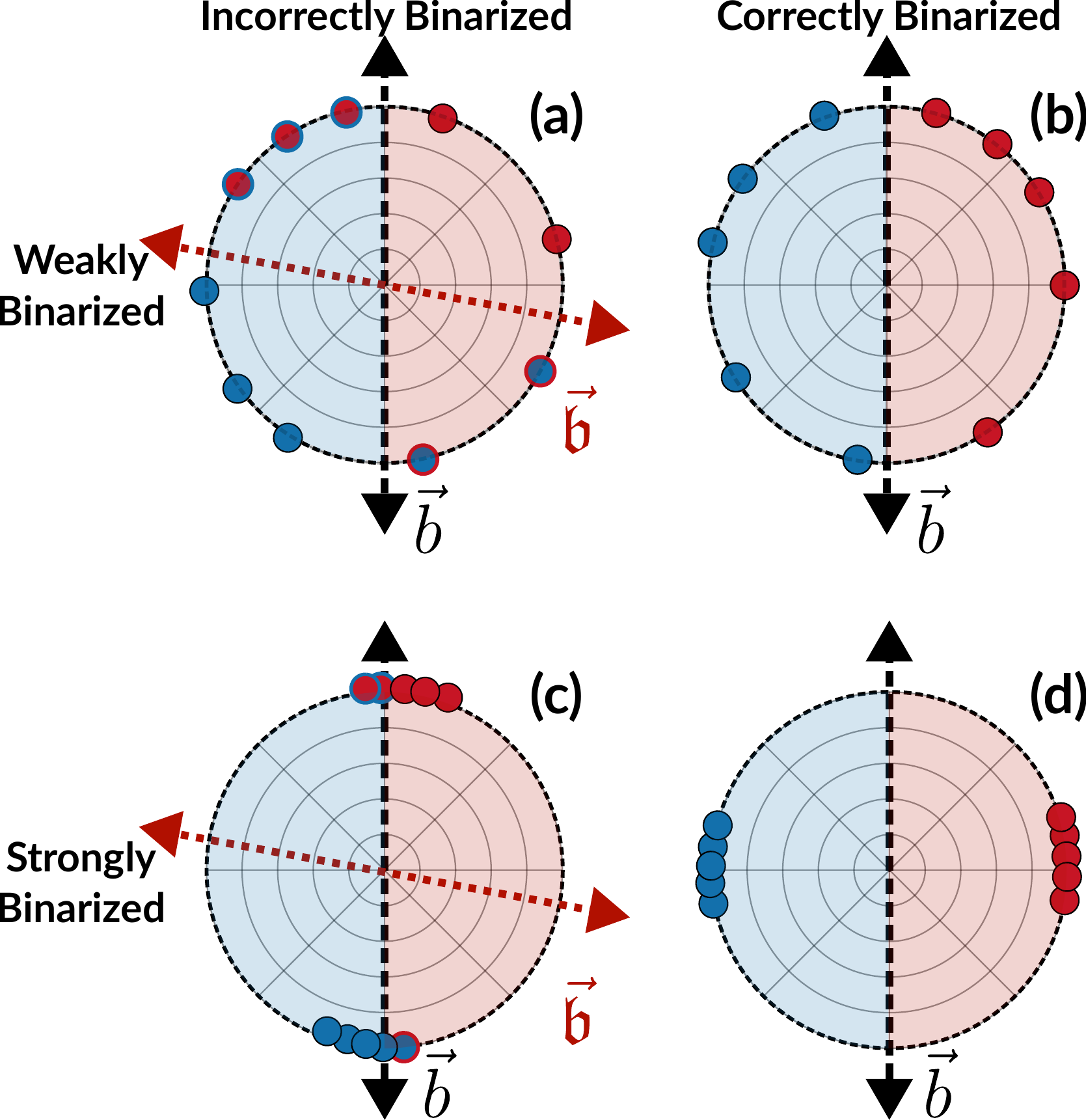}
    \caption{Example phase plots, where a system can either be strongly/weakly binarized (which corresponds the variance of the phase distribution) and correctly/incorrectly binarized (which corresponds to classification errors from axis misalignment).}
    \label{fig:bin_illustration}
\end{figure}

Recall that reading out solutions to QUBO problems from OIMs requires converting each oscillator’s continuous phase (from $0$ to $2\pi$) into discrete Ising spins ($+1$ or $-1$).  Two (somewhat independent) factors can impact the final results: \ding{172} the state of the oscillators just prior to the readout and \ding{173} the actual readout process:
\begin{enumerate}

    \item \textbf{The state:} At the end of annealing, depending on the governing dynamics, the oscillators can strongly binarize or not. Clearly, a strongly binarized system (Fig.~\ref{fig:bin_illustration}-c or d) is more robust to classification error than a weakly binarized system (Fig.~\ref{fig:bin_illustration}-a or b). 

    \item \textbf{The readout:} Mathematically, the readout step can be represented as introducing a binarization axis $\vec{b}$, which separates oscillators into two groups by their phases. For convenience, we can treat $\vec{b}$ as the imaginary axis with the associated binarization function \mbox{$f_B(z)=\text{sign}(\Re(z))$}. This axis may or may not be the binarization axis $\vec{\mathfrak{b}}$ corresponding to the lowest energy assignment, depending on the phase relationship between the oscillators and the readout circuit, leading to four cases illustrated in Fig.~\ref{fig:bin_illustration}.
\end{enumerate}
Our Wirtinger potentials have implications for both factors as we explore next.

\color{black}

Whether $\vec{\mathfrak{b}}=\vec{b}$ depends on the problem formulation. Consider a quadratic Hamiltonian $\mathcal{H}(s_1,s_2)=s_1s_2$. $\mathcal{H}$ has three main Wirtinger potentials, the 0-potential $\Hp{2}{0}(z_1,z_2)$ , the 1-potential $\Hp{2}{1}(z_1,z_2)$ , and their convex combinations, including the ``complete'' C-potential $\Hp{2}{C}(z_1,z_2)$. Their explicit forms are given by
\begin{equation}
\begin{split}
\Hp{2}{0}(z_1,z_2)&=z_1z_2 + z^*_1z^*_2 \\
\Hp{2}{1}(z_1,z_2)&=z_1^*z_2 + z_1z^*_2 \\
\Hp{2}{C}(z_1,z_2)&=0.5 (z_1z_2 + z^*_1z^*_2) \\&+ 0.5 (z_1^*z_2 + z_1z^*_2).
\end{split}
\end{equation}
Each potential gives identical values at $z_1, z_2\in\{|z|e^{i0}, |z|e^{i\pi}\}$ (typical binarization reference points). However, they do not behave identically under phase translations. Consider a global phase offset $\phi$. The 1-potential remains invariant, as

\begin{equation}
\begin{split}
    \Hp{2}{1}(e^{i\phi}z_1,e^{i\phi}z_2)&=e^{i(\phi-\phi)}z_1^*z_2 + e^{i(\phi-\phi)}z_1z^*_2\\
    &=z_1^*z_2 + z_1z^*_2=\Hp{2}{1}(z_1,z_2).
\end{split}
\end{equation}
The same is not true, however, of the 0-potential or C-potential, where
\begin{equation}
\begin{split}
    \Hp{2}{0}(e^{i\phi}z_1,e^{i\phi}z_2)&=e^{i2\phi}z_1z_2+e^{-i2\phi}z_1^*z_2^*\\
    &\neq\Hp{2}{0}(z_1,z_2)
\end{split}
\end{equation}
and
\begin{equation}
\begin{split}
    \Hp{2}{C}(e^{i\phi}z_1,e^{i\phi}z_2)&=
    0.5 e^{i2\phi}z_1z_2+0.5 e^{-i2\phi}z_1^*z_2^*\\&+ 0.5(z_1^*z_2+z_1z_2^*)\\
    &\neq\Hp{2}{C}(z_1,z_2)
\end{split}
\end{equation}
in general. For instance, for $z_1=e^{i0}$, $z_2=e^{i\pi}$, $\phi=\pi/3$, we have 
\begin{equation}
\begin{split}
    \Hp{2}{0}(e^{i\phi}z_1,e^{i\phi}z_2)&=\frac{1}{2}[\cos(\pi+\pi/3)+\cos(\pi-\pi/3)] \\&=-0.5\neq -1.0=\Hp{2}{0}(z_1,z_2).
\end{split}
\end{equation}

Investigating the Hessians provides more context. For convenience, we assume that the local dynamics confine each oscillator to unit amplitude, allowing for a phase-reduced model
\begin{equation}
\begin{split}
    \Hp{2}{0}(\theta_1, \theta_2)&=\cos(\theta_1+\theta_2)\\
    \Hp{2}{1}(\theta_1, \theta_2)&=\cos(\theta_1-\theta_2)\\
    \Hp{2}{C}(\theta_1, \theta_2)&=\frac{1}{2} \cos(\theta_1-\theta_2) + \frac{1}{2}\cos(\theta_1+\theta_2)
\end{split}
\end{equation}
with Hessians
\begin{equation}
\begin{split}
    \nabla^2\Hp{2}{0}(\theta_1, \theta_2)&=\begin{bmatrix}
        -\cos(\theta_1+\theta_2)&-\cos(\theta_1+\theta_2)\\
        -\cos(\theta_1+\theta_2)&-\cos(\theta_1+\theta_2)
    \end{bmatrix}\\
    \nabla^2\Hp{2}{1}(\theta_1, \theta_2)&=\begin{bmatrix}
        -\cos(\theta_1-\theta_2)&\cos(\theta_1-\theta_2)\\
        \cos(\theta_1-\theta_2)&-\cos(\theta_1-\theta_2)
    \end{bmatrix}\\
    \nabla^2\Hp{2}{C}(\theta_1, \theta_2)&=\frac{1}{2}\nabla^2\Hp{2}{0}(\theta_1, \theta_2)+\frac{1}{2}\nabla^2\Hp{2}{1}(\theta_1, \theta_2)
\end{split}
\end{equation}
by the linearity of the Hessian operator.

 $\nabla^2\Hp{2}{1}(\theta_1, \theta_2)$ and $\nabla^2\Hp{2}{0}(\theta_1, \theta_2)$ have eigenvalues $\{0,2\}$, however their null spaces $\mathcal{N}$ are disjoint. Specifically, $\mathcal{N}(\nabla^2\Hp{2}{0})=\text{Span}\{(1,-1)^\top\}$, $\mathcal{N}(\nabla^2\Hp{2}{1})=\text{Span}\{(1,1)^\top\}$. In other words, the 1-potential has zero curvature along the $(1,1)^\top$ vector while being strongly convex along $(1,-1)$. In contrast, the 0-potential is strongly convex along $(1,1)^\top$ while having zero curvature along $(1,-1)$. Neither potential is strongly convex in the neighborhood of their fixed points, causing each to have an uncountable number of solutions. 

However, the convex combination $\nabla^2\Hp{2}{C}(\theta_1, \theta_2)$ is full-rank and positive definite in the neighborhood of all $\{0,\pi\}$ minima, meaning it guarantees descent to the unique phase minimizer within each neighborhood. 

SHIL performs a similar function, adding the diagonal matrix
\begin{equation}
    \nabla^2 \Hp{2}{\text{SHIL}}=2\begin{bmatrix}
        \cos(2\theta_1) & 0\\
        0&\cos(2\theta_2)
    \end{bmatrix},
\end{equation}
which makes $\nabla^2\{\Hp{2}{1}+\Hp{2}{\text{SHIL}}\}$ full rank and positive definite in the neighborhood of $\{0,\pi\}$. Therefore, we have two options for binarizing quadratic Hamiltonians:
\begin{enumerate}
    \item Add $\Hp{2}{\text{SHIL}}$ to the problem Hamiltonian.
    \item Use $\Hp{2}{C}(z_1,z_2)$.
\end{enumerate}

Option 1 is typical in OIM literature~\cite{wang_oim_2019,maher_cmos-compatible_2024}. However, our analysis offers an alternative. $\Hp{2}{C}(z_1,z_2)$ or, more generally, a convex combination of the 1-potential $\Hp{2}{1}(z_1,z_2)$ and the 0-potential $\Hp{2}{0}(z_1,z_2)$ creates a strongly convex function in the neighborhood of $\{0,\pi\}$ fixed points. In the next section, we explore this possibility to induce binarization in 3-SAT problems while maintaining high-quality solutions.

\subsection{Binarization Behavior: SAT Problems}
Similar to the quadratic case, there are multiple choices to formulate a cubic oscillator model. Our empirical analysis on 3-SAT problems shows that the model defined by \eqref{eqn:improved_hopf} attains the highest solution quality, but only weakly binarizes. Conversely, replacing the terms in ~\eqref{eqn:improved_hopf} with the complete form in \eqref{eqn:h0_3} yields a new energy function which strongly binarizes but suffers from inferior optimization performance. In our experiments, this variant solved 61.8\%, 17\%, 2\%, and 0\% of the instances for the \texttt{uf-50}, \texttt{uf-75}, \texttt{uf-100}, and \texttt{uf-150} benchmarks, respectively, based on 100 independent trials per instance, which is similar to the baseline we compared with. This trade-off motivates the question: Can we design an oscillator network that simultaneously (i) achieves high solution quality, (ii) respects the Ising Hamiltonian, and (iii) yields binary spin values at convergence?

To address this, we examine the contributions of each term in the formulation \eqref{eqn:improved_hopf} independently. Linear and cubic terms inherently attain weak binarization, as their Hessians have non-zero curvature along $(1)^\top, (1,1,1)^\top$. The primary difference between the Hamiltonians lies with the quadratic terms. As we showed in the previous subsection, C-potential quadratic Hamiltonians are strongly convex in the vicinity of $\{0,\pi\}$ phase points, promoting strong binarization. In contrast, the 1-potential does not binarize, as the system is invariant to global phase shifts.

Therefore, introducing a time-varying quadratic Wirtinger potential can provide a strong binarization force while preserving the optimization performance of the oscillator network. More specifically, we can get a model with a stronger binarization tendency by selecting $\Hp{2}{C}$ for the quadratic terms as follows:
\begin{equation}\label{eqn:2}
\begin{aligned}
\mathcal{H''} &= \frac{1}{2} \sum_{i=1}^{n} h_i (z_i + z_i^*) \\
&+ \boxed{\frac{1}{4} \sum_{i=1}^{n} \sum_{j=1}^{n} J_{ij} (2 z_i z_j^* + z_i z_j + z_i^* z_j^*)} \\
&+ \frac{1}{2} \Bigl(\sum_{i=1}^{n} \sum_{j=1}^{n} \sum_{k=1}^{n} P_{ijk} z_i z_j z_k^* \\
&\quad + \sum_{i=1}^{n} \sum_{j=1}^{n} \sum_{k=1}^{n} P_{ijk} z_i z_j^* z_k^*\Bigr) + \text{constant}.
\end{aligned}
\end{equation}

It is important to note that after binarization, both Hamiltonian \eqref{eqn:improved_hopf} and \eqref{eqn:2}, as well as any combination of them, yield identical values to the discrete Ising formulation, although binarization introduces a slight modification in the energy landscape. To demonstrate the effectiveness of our binarization strategies, we evaluated four scenarios on 45 \texttt{uf100-430} 3-SAT instances:

\begin{itemize}[itemsep=-3pt, leftmargin=2em]
    \item \textbf{None:} The system evolves under Hamiltonian \eqref{eqn:improved_hopf}.
    \item \textbf{Static Potential:} The system evolves under Hamiltonian \eqref{eqn:2}.
    \item \textbf{Annealed SHIL:} The system uses Hamiltonian \eqref{eqn:improved_hopf} augmented with a SHIL signal, where the coefficient increases linearly.
    \item \textbf{Annealed Potential:} The system starts with Hamiltonian \eqref{eqn:improved_hopf} and linearly transitions to Hamiltonian \eqref{eqn:2} to achieve binarization.
\end{itemize}

\begin{figure}[h]
    \centering
    \includegraphics[width=0.45\textwidth]{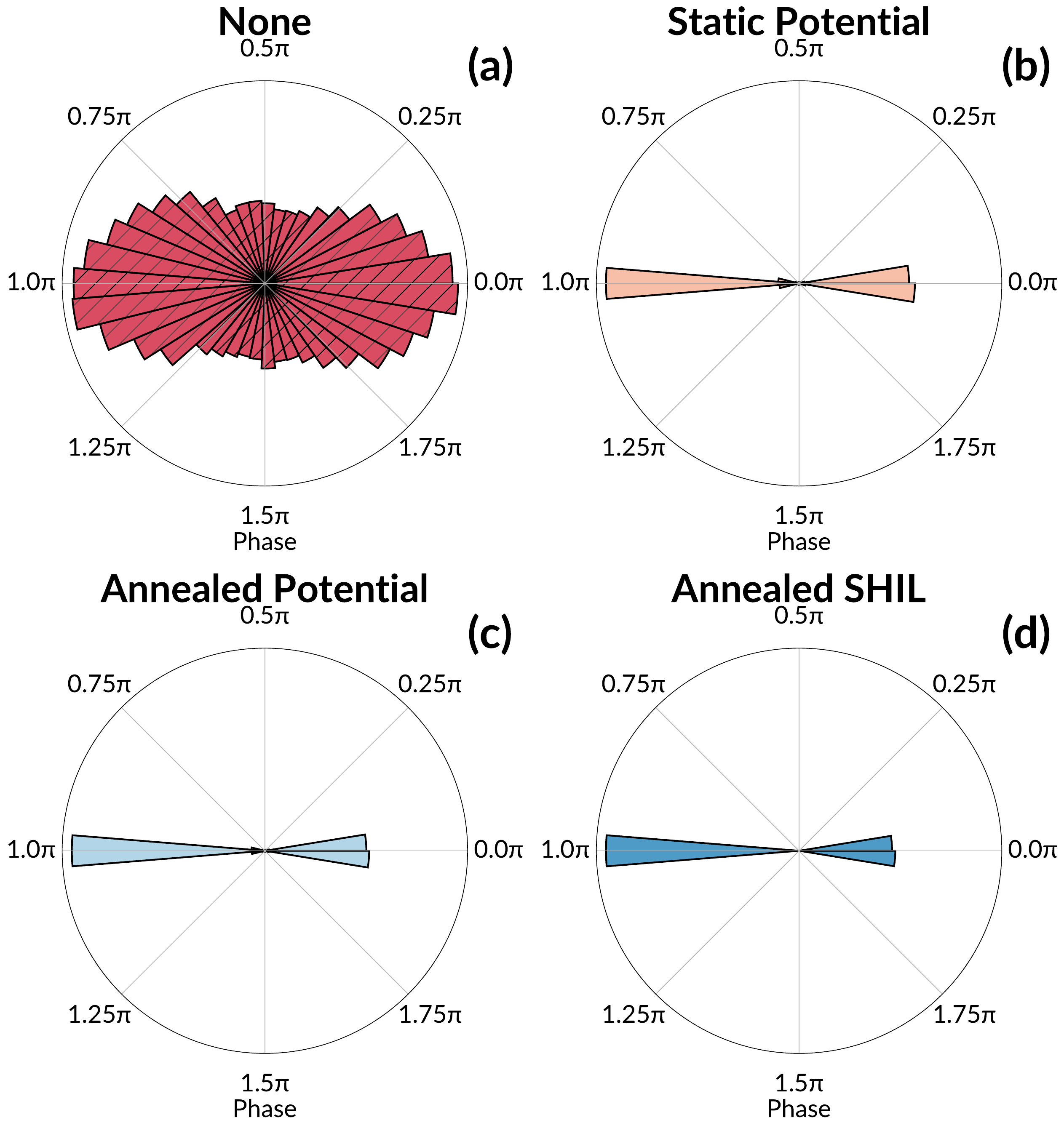}
    \caption{Polar histograms of phase values for the different binarization strategies discussed.}
    \label{fig:phase_dist}
\end{figure}

Fig.~\ref{fig:phase_dist} presents the distribution of the final oscillator phases. In the \emph{None} scenario (a), the distributions are weakly peaked at $0$ and $\pi$. However, a significant proportion of the phases are spread between the peaks, indicating little inherent binarization. In contrast, the SHIL, Static, and Annealed strategies (b)$-$(d) exhibit a strong binarization tendency. 

\begin{figure}[h]
    \centering
    \includegraphics[width=0.4\textwidth]{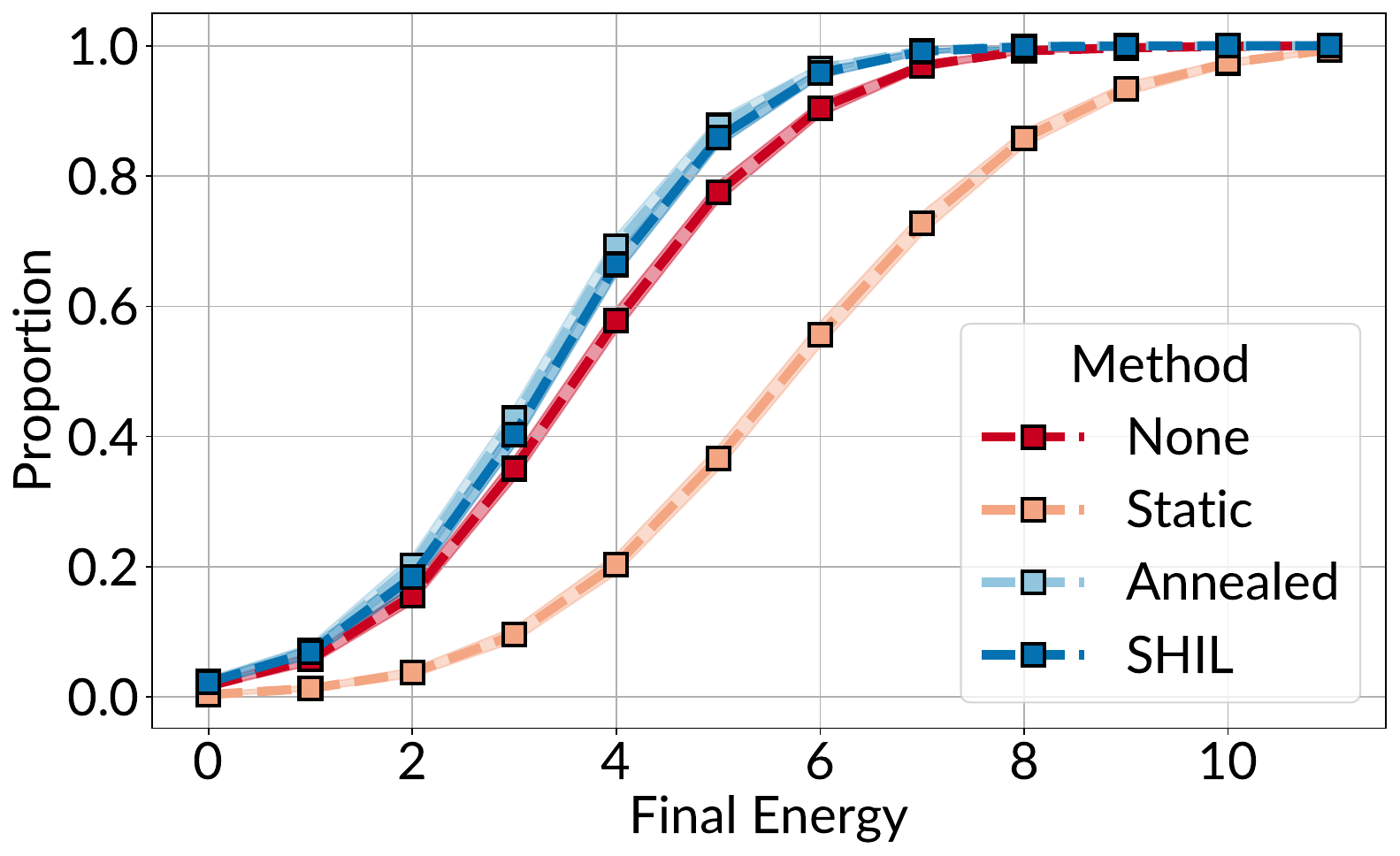}
    \caption{Cumulative distribution of final energy values for the different binarization strategies discussed. Each line represents the distribution of 45 problems from \texttt{uf100-430} with 100 trials each. 99\% confidence intervals are provided as shaded lines around each estimate, however they are typically too narrow to be visible.}
    \label{fig:unsat_cdf}
\end{figure}

Fig.~\ref{fig:unsat_cdf} shows the cumulative final energy distribution of each binarization strategy. The distributions represent 45 problems from \texttt{uf100-430} ($n=100$, $m=430$), each with 100 runs. We use bootstrap resampling with $10^4$ resamples to obtain 99\% confidence intervals (shaded area around lines); however, in most cases, they are too narrow to be easily visible. 

In lieu of final energy, a narrower figure of merit (FoM) should be just the probability of reaching the ground state (energy=0). However, we note that these are stochastic results, and one way to better see through the underlying reality is to see the distribution of all solutions. (Another way is to show the narrow FoM over more test benchmarks.) In the cumulative distribution, the higher the curve, the better. As we can see, the SHIL and Annealed strategies deliver performance that is similar to or slightly better than that achieved by the system evolving under Hamiltonian {\eqref{eqn:improved_hopf}} alone. We hypothesize that the induced binarization acts as a form of stochastic perturbation, which may help the system escape local minima during the optimization process.

\section{DISCUSSION}

Phase reduction approaches, including the Kuramoto model, fail to capture the full dynamics of oscillator-based Ising machines in real-world applications. We present a comprehensive Hopf oscillator model that explicitly incorporates amplitude dynamics, addressing this fundamental limitation. Compared to the existing baseline, our model defines a clear energy function with corresponding dynamics and achieves better performance.

Our study focuses on the theoretical soundness of phase–amplitude models for oscillator-based Ising machines. Recent studies \cite{sharma_augmenting_2023, sharma_combining_2023, salim_ski-sat_2024} demonstrate that integrating heuristic mechanisms substantially improve Ising-machine-based 3-SAT solver performance. Our model enhancements establish a solid foundation for these advances.

Incorporating amplitude dynamics introduces an additional degree of freedom. The use of a complex oscillator representation paves another way for further research. We interpret the local dynamics term $f(z_i)$ in the oscillator model (\eqref{eqn:overall_model}) not only as a mechanism to preserve physical oscillator properties but also as a means to improve the performance of the oscillator-based Ising machine. Here, we provide a perspective for future studies: oscillator-based Ising machines rely on binarization and nonlinearity to achieve better performance on MAX-CUT problems \cite{wang_oim_2019, vaidya_creating_2022}, whereas Linear Ising Machines (LIMs) like BRIM \cite{afoakwa_brim_2021} do not. By engineering the local dynamics, one can steer the oscillator-based Ising machine to mimic the dynamics of LIMs. For example, in quadratic MAX-CUT problems, the oscillator dynamics $f(z_i)$ could be designed to preserve the real part of the oscillator representation $z_i$ while adjusting the imaginary part to maintain essential oscillator properties. This approach shifts the optimization focus from system-level strategies to tuning inherent oscillator properties, although further experimental validation remains necessary.

\section{CONCLUSION}

Our work is driven by the need for a more accurate model for oscillator-based Ising machines. We observed several fundamental inconsistencies in existing phase-amplitude models. By addressing these issues, we propose a model that incorporates both phase and amplitude dynamics, features a physically meaningful Lyapunov function, and remains compatible with the widely used phase-only models in oscillator-based Ising machine implementations. Experiments show that our model outperforms existing phase-amplitude approaches in optimization performance and exhibits a correct energy minimization tendency. 

Our phase-amplitude model provides a more faithful framework for modeling oscillator-based Ising machines when amplitude dynamics cannot be neglected. This includes spintronic oscillators that naturally exhibit coupled amplitude and phase dynamics, NEMS oscillators operating in regimes where amplitude variations are significant, and emerging photonic systems. By providing a rigorous mathematical foundation that captures these dynamics, our model enables more accurate simulations and predictions of system behavior, which is crucial for the design and optimization of OIMs. Accurate modeling further enhances practical applications, including portfolio optimization, drug discovery, wireless communication, and a wide range of other combinatorial optimization problems.

Furthermore, we propose a method to generalize the model for constructing higher-order Ising machines and an implementable approach for extracting solutions from the oscillators. These findings contribute to both the theoretical understanding and practical realization of oscillator-based Ising machines, offering valuable directions for future research and development in this promising field.

\section*{ACKNOWLEDGMENTS}
This work was supported in part by NSF under Awards No. 2231036 and No. 2233378, and by DARPA under contract No. FA8650-23-C-7312.

\section*{DATA AVAILABILITY}
The data that support the findings of this article are
openly available \cite{ACAL_prapplied-oscillator-data}.

\section*{APPENDIX}
\appendix
\newcommand{\myhl}[1]{\colorbox{yellow}{#1}}

\section{Example: Mapping a 3-SAT problem to Ising Hamiltonian}\label{appdx:sat}

We illustrate the mapping on the single clause
$$
C = (x_1 \lor \neg x_2 \lor x_3).
$$

Our goal is to construct a Hamiltonian $H_C$ for this clause. When $H_C$ is minimized to 0, the clause is satisfied; otherwise, it contributes 1 to the total energy.

\medskip
\noindent\textbf{Mapping Method:}
Each Boolean $x_i\in\{0,1\}$ is represented by an Ising spin $s_i\in\{+1,-1\}$, with
$$
s_i = +1 \iff x_i = \text{true}, 
\quad
s_i = -1 \iff x_i = \text{false}.
$$

\medskip
\noindent\textbf{Step 1: Literal energy terms.} 
For each literal $\ell$, define
$$
f(\ell) =
\begin{cases}
1 - s_i, & \ell = x_i,\\
1 + s_i, & \ell = \neg x_i.
\end{cases}
$$
In our clause,
$$
f(x_1) = 1 - s_1,\quad f(\neg x_2) = 1 + s_2,\quad f(x_3) = 1 - s_3,
$$
according to the mapping method, $f(\ell)=0$ when the literal is true, and $f(\ell)=2$ otherwise.

\medskip
\noindent\textbf{Step 2: Clause energy by multiplication.} 
To ensure the energy is nonzero only when all three literals fail, we take the product:
$$
(1 - s_1)\,(1 + s_2)\,(1 - s_3)
=
\begin{cases}
0, & \text{if at least one literal is true},\\
8, & \text{if all three literals are false}.
\end{cases}
$$

\medskip
\noindent\textbf{Step 3: Normalization.}
Dividing by 8 yields the clause Hamiltonian
$$
H_C
= \frac{(1 - s_1)\,(1 + s_2)\,(1 - s_3)}{8}
=
\begin{cases}
0, & \text{clause satisfied},\\
1, & \text{clause unsatisfied}.
\end{cases}
$$

\medskip
Finally, for a full 3-SAT problem with clauses $C_1,\dots,C_M$, we sum these terms:
$$
\mathcal{H} = \sum_{j=1}^{M} H_{C_j},
$$
so $\mathcal{H}$ counts exactly the number of unsatisfied clauses.

\section{Wirtinger Calculus for Non-Holomorphic Hamiltonians}\label{appdx:wirtinger}

Let $z=x+iy$ be a complex $n$-dimensional vector, where $x,y\in\R^n$. When the Hamiltonian $\mathcal{H}$ depends on both $z=x+iy$ and its complex conjugate $z^*=x-iy$, it is typically not holomorphic and the usual derivative does not exist. Wirtinger calculus treats $z$ and $z^*$ as independent variables. For any real-valued $\mathcal{H}(z,z^*)$, Wirtinger partial derivatives are defined as
$$
\pdv{\mathcal{H}}{z_i}\triangleq \frac{1}{2}\left(\pdv{\mathcal{H}}{x_i}-i\pdv{\mathcal{H}}{y_i}\right),
$$

$$
\pdv{\mathcal{H}}{z_i^*}\triangleq \frac{1}{2}\left(\pdv{\mathcal{H}}{x_i}+i\pdv{\mathcal{H}}{y_i}\right).
$$  

As a consequence of the definitions above, we have 
\begin{align*}
\pdv{\mathcal{H}}{z_i}=\left(\pdv{\mathcal{H}}{z_i^*}\right)^*
\\
\pdv{\mathcal{H}}{z_i^*}=\left(\pdv{\mathcal{H}}{z_i}\right)^*.
\end{align*}
While the formal definitions are expressed using $x$ and $y$, the Wirtinger partial derivatives behave as expected for a function defined in terms of $z$ and $z^*$. For example, one can easily verify that $\pdv{}{z}[zz^*]=z^*$ and $\pdv{}{z^*}[zz^*]=z$.

Additionally, a function $\mathcal{H}$ is holomorphic if and only if $\partial \mathcal{H}/\partial z_i^*=0$ for all $z_i$.
Intuitively, Wirtinger calculus treats $\mathcal{H}$ as a function over $\R^{2n}$ instead of $\C^n$, using appropriately defined isomorphisms to transform between $\mathcal{H}(x,y):\R^{2n}\to\R$, $ \mathcal{H}(z,z^*):\C^{2n}\to\R$ and $\mathcal{H}(z):\C^{n}\to\R$. From this perspective, typical objects from multivariate calculus (gradients, Hessians, $\dots$) have a natural extension to real-valued, non-holomorphic functions. Additional details can be found in the excellent introduction to the subject by Kreutz-Delgado~\cite{kreutz-delgado_complex_2009}. Crucial for our purposes is the definition of the Wirtinger gradient
\[
    \nabla_{z} \mathcal{H}\triangleq \left(\pdv{\mathcal{H}}{z}\right)^\dagger,
\]
where $\pdv{\mathcal{H}}{\bm{z}}\in\C^{2n}$ is the vector
\[
\left(\pdv{\mathcal{H}}{z_1},\dots,\pdv{\mathcal{H}}{z_n},\pdv{\mathcal{H}}{z_1^*},\dots,\pdv{\mathcal{H}}{z_n^*}\right)^\top
\] and $(\cdot)^\dagger$ denotes a conjugate transpose. Since Wirtinger partials have the bijective relationship $\pdv{\mathcal{H}}{z_\ell}=\left(\pdv{\mathcal{H}}{z_\ell^*}\right)^*$, we focus only on the non-conjugated components.

\medskip
\noindent\textbf{Example:} Consider an energy function 
$
\mathcal{H}(z,z^*) = z\,z^*.
$
Treating $z^*$ as constant when differentiating with respect to $z$ (and vice versa), we obtain
$$
\frac{\partial \mathcal{H}}{\partial z}
= z^*,
\qquad
\frac{\partial \mathcal{H}}{\partial z^*}
= z.
$$
Since $\partial \mathcal{H}/\partial z^*=z\neq0$, $\mathcal{H}$ is non-holomorphic.

\medskip
\noindent\textbf{Energy Minimization via Gradient Descent:}  
For a real-valued Hamiltonian $\mathcal{H}(z,z^*)$, a small update $\Delta z$ produces
$$
\Delta \mathcal{H} \approx \frac{\partial \mathcal{H}}{\partial z}\,\Delta z
+
\frac{\partial \mathcal{H}}{\partial z^*}\,\Delta z^*.
$$
Recall that $\pdv{\mathcal{H}}{z^*}=\left(\pdv{\mathcal{H}}{z}\right)^*$ and vice versa.
Therefore, to minimize $\Delta\mathcal{H}$, it suffices to choose
\begin{equation}
    \Delta z \propto -\pdv{\mathcal{H}}{z^*}
\end{equation}
which produces
\[\Delta \mathcal{H} \propto-2\left\lVert\pdv{\mathcal{H}}{z^*}\right\rVert^2.\]

Choosing a ``step size'' of $\eta$, we therefore have a well-defined gradient descent scheme:
$$
z^{k+1} \leftarrow z^{k}  -\eta\,\frac{\partial \mathcal{H}}{\partial z^*}.
$$
In the previous example $\partial \mathcal{H}/\partial z^*=z$, so iterating 
$z^{k+1}\leftarrow z^{k}-\eta\,z^{k}$
drives $z\to0$, the minimizer of the energy $\mathcal{H}=|z|^2$.

\medskip
\section{General PUBO Wirtinger Potentials}\label{appdx:wirtinger_advanced}
As defined in Sec.~\ref{sec:3_bridge}, we can express the ``complete'' $k^\textrm{th}$ order PUBO Hamiltonian $\Hp{k}{C}:\C^n\to\R$ as
$$
    \Hp{k}{C}(z)=\frac{1}{\lfloor k/2\rfloor}\sum_{p=1}^{\lfloor k/2\rfloor}\Hp{k}{p}(z),
$$
where each $p$-potential has $m_p=2\binom{k}{p}$ terms and each $\Hp{k}{p}:\C^n\to\R$ has the form
\begin{align*}
\Hp{k}{p}(z)=\frac{1}{2^k}\sum_{j=1}^{m_p}c_{p,j}\left(y^{(1)}_{p,j}\dots y^{(p)}_{p,j}(y^{(p+1)}_{p,j})^*\dots (y^{(k)}_{p,j})^*\right.\\
+\left.(y^{(1)}_{p,j})^*\dots (y^{(p)}_{p,j})^*y^{(p+1)}_{p,j}\dots y^{(k)}_{p,j}
\right) ,
\end{align*}
where $p$ denotes the number of non-conjugated terms in the first product and each term $y^{(q)}_{p,j}$ refers to some complex variable $z_\ell$ appearing as the $q^{\text{th}}$ term in the $j^\textrm{th}$ product.

\bibliographystyle{unsrt}
\bibliography{ACAL, additional} 
\end{document}